\documentclass[12pt]{iopart}
\pdfoutput=1

\usepackage{iopams}  
\usepackage{graphicx}
\usepackage{color}
\begin{document}

\title[Dynamic heterogeneity lengths in two dimensions]
{Dynamic heterogeneity in two-dimensional supercooled liquids: comparison of bond-breaking and bond-orientational
correlations}

\author{Elijah Flenner \& Grzegorz Szamel}

\address{Department of Chemistry, Colorado State University, Fort Collins, CO 80523, USA}
\vspace{10pt}
\begin{indented}
\item[]\today
\end{indented}

\begin{abstract}
We compare the spatial correlations of bond-breaking events and bond-orientational relaxation in a model two-dimensional 
liquid undergoing Newtonian dynamics. We find that the relaxation time of 
the bond-breaking correlation function
is much longer than the relaxation time of the bond-orientational correlation function
and self-intermediate scattering function. However, the relaxation time of the bond-orientational
correlation function increases faster with decreasing temperature than the relaxation time
of the bond-breaking correlation function and the self-intermediate scattering function. 
Moreover, the dynamic correlation length that characterizes the size of correlated bond-orientational
relaxation grows faster with decreasing temperature than the dynamic correlation length that
characterizes the size of correlated bond-breaking events.  We also examine the
ensemble-dependent and ensemble-independent dynamic susceptibilities 
for both bond-breaking correlations and bond-orientational correlations. We find that for both correlations, 
the ensemble-dependent and ensemble-independent susceptibilities exhibit a maximum 
at nearly the same time, and this maximum occurs at a time slightly shorter than the peak position
of the dynamic correlation length. 
\end{abstract}

%
%
\submitto{\JSTAT}
%
%
%

\section{Introduction}
 One of the most studied features of glassy dynamics is the existence of transient 
 domains in which particles move in a correlated fashion, either much farther 
 or much less than expected from a Gaussian distribution of displacements \cite{LudoBook}. 
 A connection between this dynamic heterogeneity and the structure was established in studies of the 
 isoconfigurational ensemble in two-dimensions \cite{Widmer2004}. In these studies the same initial 
 structure was used in many simulation runs, with the velocities chosen randomly from a 
 Boltzmann distribution. It was found that there were regions where the
 particles were more likely to move farther than in other regions, and thus
 some aspects of the local structure were associated with the domains of heterogeneous dynamics.  
 The exact nature of these aspects of the local structure is, however, not clear. 

 Recently it has been shown that there is a large difference in 
 the growth of the characteristic size of dynamically heterogeneous regions upon supercooling 
 in two and three dimensions if one defines dynamic
 heterogeneity through particle displacements \cite{Flenner2015}. This difference 
 is mirrored in strong finite size effects in the quantities that probe translational motion of the
 tagged particle, e.g.\ the self-intermediate scattering function
 and the mean-square displacement, in two-dimensional glass forming liquids. 
 Despite the difference in the tagged particle dynamics, there may be other
 dynamic observables that exhibit similar behavior in two and three dimensions.
 One can gain insight into potential observables that might exhibit similar behavior when
 one considers the dimensional dependence of the formation of ordered solids \cite{Strandburg1988}.
 Similar to the glass transition, this freezing transition is also a transition from a liquid state where 
 the shear modulus is zero, to a state with non-zero shear modulus. 
 In three dimensions long-range translational and rotational order 
 emerges when an ordered solid forms, but in two dimensions 
 only long-range rotational order emerges when an ordered solid forms. 
 Thus, in two-dimensional ordered solids, bond-orientational
 order should be monitored to look for the emergence of an elastic solid. When 
 comparing the glass transition in two and three dimensions one may want to examine 
 quantities that are expected to be similar in studies of the formation of two-dimensional and three-dimensional 
 ordered solids, and this leads to examining correlations between bonds connecting neighboring particles. 
 
 Yamamoto and Onuki \cite{Yamamoto1998, Yamamoto2000} examined spatial correlations of bond-breaking events in 
 a two-dimensional glassy binary mixture. 
 They found large regions in which bond-breaking events were correlated
 by examining four-point structure factors in which the so-called weight function \cite{FlennerSzamelJPCB2015} 
 was determined by bond-breaking events. 
 Subsequently, Shiba, Kawasaki, and Onuki \cite{Shiba2012} compared bond-breaking correlations
 and mobility correlations in two-dimensional and three-dimensional systems. 
 They concluded that the vibrational modes made a significant contribution to 
 mobility correlations in two-dimensional liquids, but the contribution from vibrational 
 modes was much smaller in three dimensions. Furthermore, the vibrational modes
 did not significantly contribute to the bond-breaking correlation function in
 two or three dimensions. These results suggest that to compare supercooled dynamics in two 
 and three dimensions one should examine bond-breaking events.
 
 Such a study has recently been performed by Shiba, Kawasaki, and Kim \cite{Shiba2015}. 
 They examined mobility correlations and bond-breaking correlations
 in two and three dimensions. They found that the characteristic lengths determined from 
 bond-breaking and mobility correlations are comparable in three dimensions, but the mobility correlations are
 much more spatially extended than the bond breaking correlations in two dimensions. Furthermore, there
 were large finite size effects in the mobility correlations in two dimensions, but no
 finite size effects were seen for bond-breaking correlations in two dimensions. There were 
 no reported finite size effects in three dimensions for mobility or bond-breaking 
 correlations. 
 
 Another observable that may have similar behavior in two- 
 and three-dimensional supercooled liquids is bond-orientational correlations. 
 We showed earlier that bond-orientational
 relaxation times are decoupled from translational relaxation times in two dimensions \cite{Flenner2015}. Specifically, 
 bond-orientational relaxation times grow faster with decreasing temperature than translational relaxation times. 
 This is analogous to the decoupling of the bond-orientational and translational ordering in ordered
 two-dimensional solids.  
 However, there has been no detailed comparison of the growth of the spatial correlations of bond-breaking events and
 bond-orientational relaxation in two-dimensional supercooled liquids. 
 Here we examine spatial correlations of bond-breaking events and bond-orientational relaxation in a model 
 two-dimensional supercooled liquid.  
 
 \section{Simulations}
 
 We simulated a 65:35 binary mixture of Lennard-Jones particles in two dimensions. 
 The interaction potential is given by
 \begin{equation}
 V_{\alpha \beta}(r_{ij}) = 4 \epsilon_{\alpha \beta} \left[
 \left(\frac{\sigma_{\alpha \beta}}{r_{ij}}\right)^{12} - \left(\frac{\sigma_{\alpha \beta}}{r_{ij}}\right)^6 \right],
 \end{equation}
 where $r_{ij}$ is the distance between particle $i$ and $j$, and $\alpha$ and $\beta$ denotes the
 type of particle. We denote the majority species as the $A$ particles and the minority species as the $B$ particles; 
 the potential parameters are given by $\epsilon_{BB} = 0.5\epsilon_{AA}$, $\epsilon_{AB} = 1.5 \epsilon_{AA}$, 
 $\sigma_{BB} = 0.88\sigma_{AA}$, and $\sigma_{AB} = 0.8\sigma_{AA}$. 
 We present results in reduced units, where the unit of 
 length is $\sigma_{AA}$, the unit of temperature is $k_B/\epsilon_{AA}$, and the unit of
 time is $\sqrt{m \sigma_{AA}^2/\epsilon_{AA}}$. The mass for both species are the same,
 and the number density is $\rho = 1.2$.  
 We simulated $N=250,000$ particles at $T=1.0, 0.8, 0.7, 0.6$, and 0.5. 
 For $T=0.45$ we simulated $10\, 000$, $250\, 000$ and 4 million particle systems. 
 While no finite size effects where found for the bond-orientational
 and bond-breaking correlation functions, there are finite size effects for the self-intermediate
 scattering function for $T=0.45$; the 4 million particle system is needed to remove
 those finite size effects \cite{Flenner2015}.  
 Due to the large bond-breaking relaxation times, the simulations
 were much longer than in previous work: we ran the production runs for at least $50 \tau_B$
 for $T \ge 0.5$ where $\tau_B$ is defined in Section \ref{relaxation}.  
 At the lowest temperature, $T=0.45$, the
 production runs were 20 times longer than the longest runs used in the previous work. 
 For every temperature we ran four productions runs. 
 The simulations were run in the NVT ensemble with a Nos\'e-Hoover thermostat using HOOMD-blue \cite{HOOMD}.
 Importantly, we note that these simulations evolve according to Newtonian dynamics. For two-dimensional systems,
 it has been shown that the underlying dynamics influences the characteristic size of dynamically heterogeneous regions
 if one defines dynamic heterogeneity through particle displacements \cite{Flenner2015}. 
 
 \section{Relaxation functions and their characteristic times}
 \label{relaxation}
 We begin by examining correlation functions related to particle displacements,
 bond-orientational relaxation, and the lifetime of bonds between neighboring particles. 
 First, we study the self-intermediate scattering function
 \begin{equation}
 F_s(q;t) = \frac{1}{N} \left< \sum_n e^{i \mathbf{q} \cdot (\mathbf{r}_n(t) - \mathbf{r}_n(0))} \right>,
 \end{equation}
 where $q = \left| \mathbf{q} \right| = 6.28$ is chosen to facilitate comparison with 
 previous work \cite{Flenner2015,Hocky2013}. 
 \begin{figure}
 \centerline{\includegraphics[width=12cm]{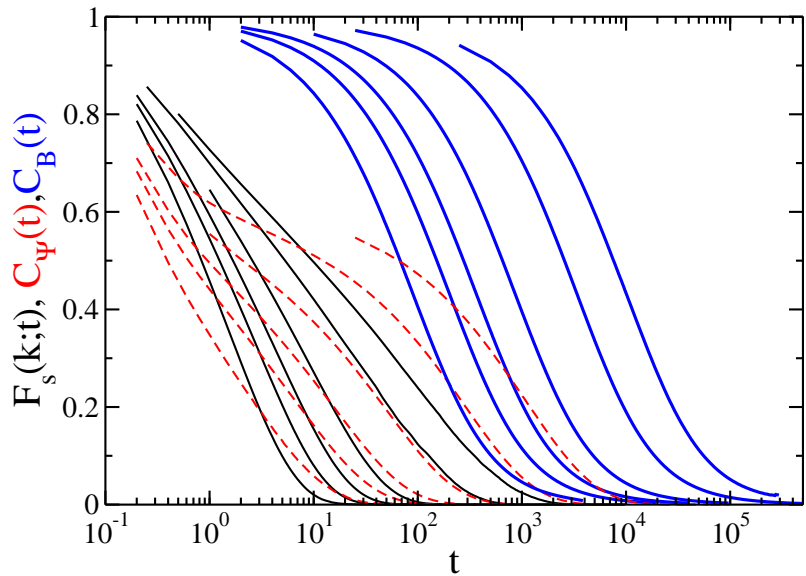}}
 \caption{\label{correlation}The self-intermediate scattering function $F_s(k;t)$ (solid black lines), 
 the bond-orientational correlation $C_{\Psi}(t)$ (dashed red lines), and the bond-breaking
 correlation function $C_B(t)$ (thick blue lines) for $T=1.0, 0.8, 0.7, 0.6, 0.5$, and 0.45 listed
 from left to right. Results are shown for 250,000 particles systems except for $F_s(k;t)$ 
 at $T=0.45$, where a 4 million particle system is needed to remove finite size effects.}
 \end{figure}
 
 Bond-orientational relaxation is quantified by the bond-angle correlation function 
 \begin{equation}
 C_{\Psi}(t) = 
 \frac{\left< \sum_n \Psi_6^n(t) \left[ \Psi_6^n(0)\right]^* \right>}{\left< \sum_n \left| \Psi_6^n(0) \right|^2 \right>},
 \end{equation}  
 where $\Psi_6^n(t) = (N_b^n)^{-1} \sum_m e^{i 6 \theta_{nm}(t)}$ and $^*$ denotes the complex conjugate. 
 The sum over $m$ is over the neighbors of particle $n$
 determined through Voronoi tessellation and the sum over $n$ is over all particles in the system. 
 $N_b^n$ is the number of neighbors of particle $n$, and
 $\theta_{nm}$ is the angle made by the bond between particle $n$ and $m$ with respect to an 
 arbitrary axis. 
 
 To examine bond-breaking events, we first define bonded particles. To be consistent with previous
 work \cite{Yamamoto1998,Yamamoto2000,Shiba2012,Shiba2015}, we define particles 
 of type $\alpha$ and $\beta$ as bonded if 
 $\left| \mathbf{r}_n^{\alpha} - \mathbf{r}_m^{\beta} \right| < 1.15 \sigma_{\alpha \beta}$. 
 The bond is broken at a later time if 
 $\left| \mathbf{r}_n^{\alpha} - \mathbf{r}_m^{\beta} \right| > 1.5 \sigma_{\alpha\beta}$.
 We define a function $Z_n(t)$ which is the number of broken bonds of particle 
 $n$ at a time $t$. Therefore $Z_n(t=0)$ is zero and $Z_n(t=\infty)$ is the number 
 of bonds particle $n$ had at $t=0$.  
 The bond-breaking correlation function is defined as
 \begin{equation}
 C_B(t) = 1 - \frac{N^{-1} \left< \sum_n Z_n(t)\right>}{\left<N_B\right>},
 \end{equation}
 where $\left<N_B\right> = N^{-1} \left< \sum_n Z_n(\infty)\right>$ is the average number of bonds per particle. 
 
 Shown in Fig.~\ref{correlation} are $F_s(t)$ (solid black lines), $C_{\Psi}(t)$ (dashed red lines),
 and $C_B(t)$ (thick blue lines) as a function of time for $T=1.0, 0.8, 0.7, 0.6,$ 0.5, and 0.45 listed from
 left to right. 
 As noted in previous work \cite{Flenner2015}, there is no plateau in the self-intermediate scattering
 function if the simulated system is large enough, which is in contrast 
 to simulations of three-dimensional glass forming systems. However, a plateau does begin to emerge 
 in the bond-orientational correlation function $C_{\Psi}(t)$.  Furthermore, 
 it can be seen in Fig.~\ref{correlation} that the decay time of $C_{\Psi}(t)$ 
 increases faster with decreasing temperature than $F_s(k;t)$. The characteristic 
 decay time of $C_B(t)$ is longer then that of either $F_s(k;t)$ or $C_{\Psi}(t)$ at each
 temperature. While the particles are moving translationally and even the orientations of
 the bonds between neighboring particles relax, the particles have not lost their neighbors. 
 \begin{figure}
 \centerline{\includegraphics[width=12cm]{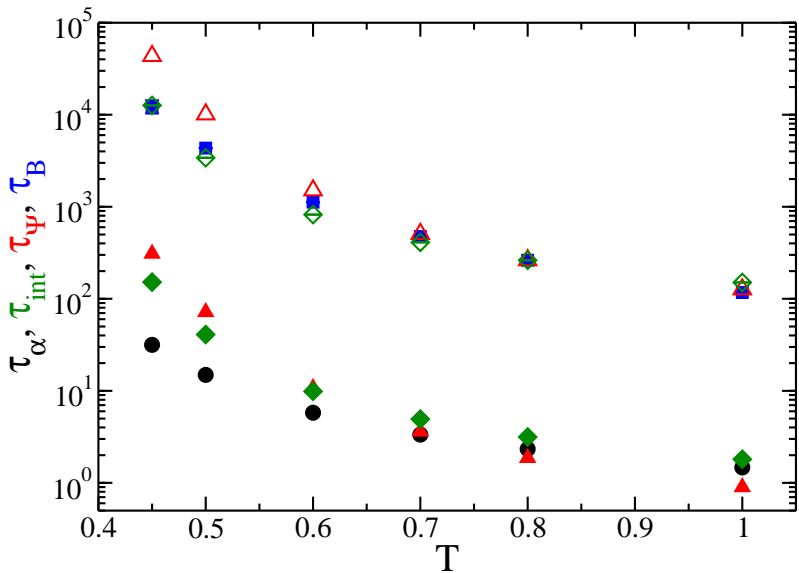}}
 \caption{\label{tau}The alpha-relaxation time $\tau_\alpha$ (black circles),
 the integrated relaxation time $\tau_{\mathrm{int}}$ (green diamonds), 
 the bond-orientational relaxation time $\tau_\Psi$ (solid red triangles), and
 the bond-breaking relaxation time $\tau_B$ (blue squares) as a function
 of temperature $T$. The open 
 red triangles are the bond-orientational relaxation times scaled
 so that the $\tau_\Psi$ equals $\tau_B$ at $T=0.8$. The open green diamonds is the integrated 
 relaxation time scaled so that equal $\tau_{\mathrm{int}}$ equals $\tau_B$ at $T=0.8$.}
 \end{figure}
 
 To quantify the decay time we define the relaxation times $\tau_\alpha$, $\tau_\Psi$, and 
 $\tau_B$ through $F_s(k;\tau_\alpha) = e^{-1}$, $C_{\Psi}(\tau_\Psi) = e^{-1}$,
 and $C_B(\tau_B) = e^{-1}$. These relaxation times are shown in Fig.~\ref{tau} as a 
 function of temperature. The relaxation time $\tau_\alpha$ grows the slowest with decreasing
 temperature. The bond-breaking relaxation times $\tau_B$ is larger than the 
 other two relaxation times over this temperature range. However, $\tau_\Psi$ grows faster 
 with decreasing temperature than either $\tau_\alpha$ or $\tau_B$. To demonstrate this faster growth, 
 we show as open red triangles $a \tau_\Psi$ where $a$ is the ratio $\tau_B/\tau_\Psi$ 
 at $T=0.8$. 
 
 We note that we defined $\tau_\alpha$ in the standard way, but this definition is 
 not the only way to determine a relaxation time. Another method would be to examine
 an integrated relaxation time $\tau_{\mathrm{int}} = \int_0^\infty F_s(k;t) dt$, which 
 would be equal to $\tau_\alpha$ as defined above if $F_s(k;t)$ decayed exponentially and
 proportional to $\tau_\alpha$ if time-temperature superposition holds. However, time-temperature
 superposition does not hold for $F_s(k;t)$ for the two-dimensional glass former, so we also examined  
 $\tau_{\mathrm{int}}$, which is shown as filled green diamonds in Fig.~\ref{tau}. 

 We find that $\tau_{\mathrm{int}}$ grows with decreasing temperature at the same rate 
 as $\tau_B$, which we demonstrate by multiplying $\tau_{\mathrm{int}}$ by a 
 constant $b$ such that $b \tau_{\mathrm{int}}$ is equal to $\tau_B$ at $T=0.8$, and
 this scaling is shown as the open green triangles in Fig.~\ref{tau}. 

\begin{figure}
 \centerline{\includegraphics[width=12cm]{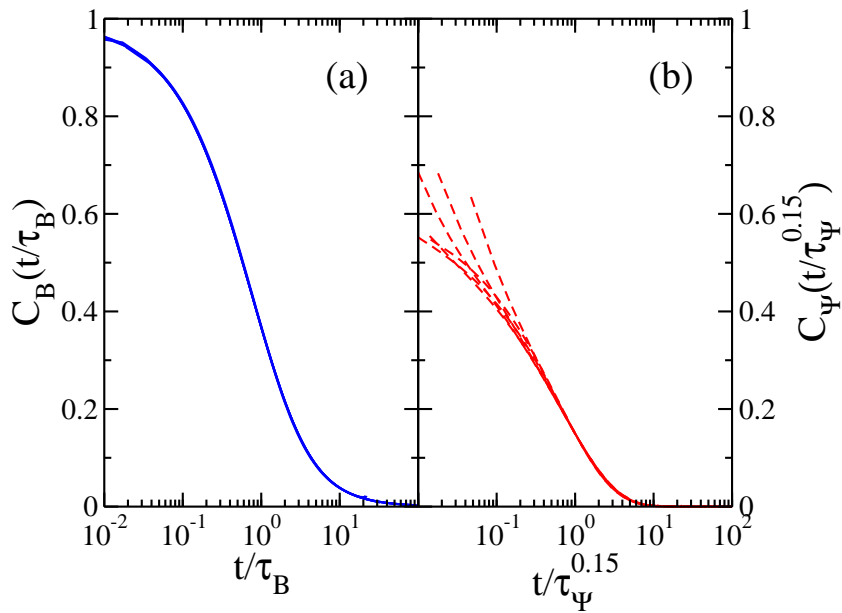}}
 \caption{\label{timetemp}
 (a)  The bond-breaking correlation function $C_B(t)$ where the time axis is rescaled by $\tau_B$.
 (b) The bond-orientational relaxation function $C_\Psi(t)$ 
 where the time axis is rescaled by $\tau_\Psi^{0.15}$ where $\tau_\Psi^{0.15}$ 
 is defined through $C_\Psi(\tau_\Psi^{0.15}) = 0.15$. 
 Time-temperature superposition holds very well for the whole time dependence of  $C_B(t)$ and for
 the final decay of $C_\Psi(t)$. In both (a) and (b) $T=1.0$, 0.8, 0.7, 0.6, 0.5, and 0.45 are shown.}
 \end{figure}

 We note that
 time-temperature superposition holds very well for $C_B(t)$, and it also holds for 
 the final decay of $C_\Psi(t)$, \textit{i.e.} for $C_\Psi(t) \le 0.2$, see Fig.~\ref{timetemp}.  
 We checked whether our results for the temperature
 dependence of the bond orientational relaxation time depend on our definition of $\tau_\Psi$. 
 If we define a relaxation time $\tau_\Psi^c$ through
 $C_\Psi(\tau_\Psi^c) = c$ with $c \le e^{-1}$, the choice of 
 $c$ does not alter our conclusions. Specifically, $\tau_\Psi^c$ with $c \le e^{-1}$ 
 increases faster with decreasing temperature than $\tau_\alpha$, $\tau_{\mathrm{int}}$,
 and $\tau_B$.  

Finally, a visual inspection of the bond-breaking and bond-orientational correlation functions suggests that 
the bond-breaking relaxation is less stretched that the bond-orientational one. 
To quantify the temperature dependence of the shape of these correlation functions we fit $C_B(t)$ and 
$C_\Psi(t)$ to a stretched exponential $A\exp(-(t/\tau)^\beta)$. We found that the 
the stretching exponent $\beta$ was independent of temperature for $C_B(t)$,
and was $0.673 \pm 0.004$. The stretching exponent for
$C_\Psi(t)$ was $\beta = 0.651 \pm 0.006$ at
$T=1.0$ and $\beta \approx 0.62 \pm 0.01$ for $0.45 \le T \le 0.8$. 
 
 \section{Dynamic Correlation Lengths}\label{lengths}
 It has been suggested that the 
 characteristic size of clusters of correlated relaxation events can
 define a dynamic correlation length that accompanies the glass transition. 
 Here we examine correlations between bond-breaking events and correlations of bond-orientational relaxation.
 To find 
 the dynamic correlation length it has become standard practice to define a four-point structure factor 
 \begin{equation}
 S_4^w(q;t) =  \frac{1}{N} \left< \sum_{nm} w_n(t) w_m^*(t) e^{i \mathbf{q} \cdot [\mathbf{r}_n(0) - \mathbf{r}_m(0)]} \right>,
 \end{equation}
 and determine a characteristic size of mobile or immobile clusters from the small $q$ 
 behavior of $S_4^w(q;t)$ \cite{LudoBook}. Importantly, the four point structure factor involves a time dependent weight 
 function $w_n(t)$ that is sensitive to the microscopic dynamic events associated with particle $n$. Thus, the weight
 function determines the dynamic events whose spatial correlations the four-point structure factor measures.  
 
 In previous work we examined the characteristic size of 
 clusters of slow particles \cite{Flenner2015}. To this end we used a weight function that selects slow particles, which we
 defined as particles that moved less than $a=0.22$, 
 $\Theta(a-|\mathbf{r}_n(t) - \mathbf{r}_n(0)|)$ where $\Theta$ is Heaviside's step function. 
 We found that the characteristic size of clusters of particles which are slow on the time scale of the $\alpha$
 relaxation time, $\tau_\alpha$, 
 grows linearly with $\tau_\alpha$ for two-dimensional systems evolving with Newtonian dynamics
 \cite{Flenner2015}. This was in contrast with the $\ln(\tau_\alpha)$ growth
 that was observed for fragile glass formers in three dimensions \cite{Flenner2014,Flenner2011}.   
 Here we examine correlations of bond-breaking events. To this end we use $Z_n(t)$ as the weight function $w_n(t)$.
 We also study correlated bond-orientational relaxation by using 
 $\Psi_6^n(t) [\Psi_6^n(0)]^*$ as the weight function. To determine 
 the dynamic correlation length $\xi_{w}(t)$ we fit 
 $S_4^{w}(q;t)$ to an Ornstein-Zernicke form, 
 $S_4^{w}(0;t)/[1+(q \xi_{w}(t))^2]$, for small $q$. We denote dynamic correlation lengths
 characterizing the spatial correlations of bond-breaking events as $\xi_B(t)$ and dynamic correlation lengths 
 measuring the spatial correlations of bond-orientational relaxation as $\xi_\Psi(t)$. For a comparison, we also
 present the dynamic correlation lengths characterizing the size of clusters of slow particles, which are
 denoted as $\xi_O(t)$.

\begin{figure}
\centerline{\includegraphics[width=12cm]{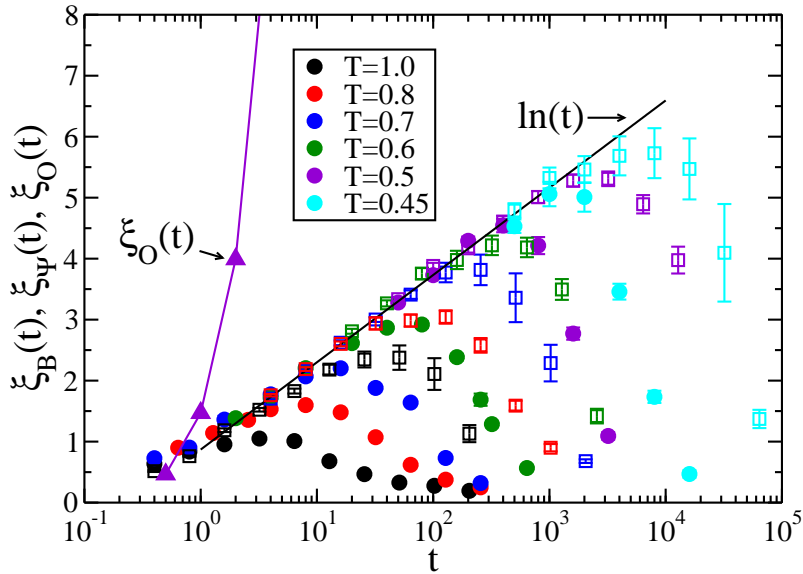}}
\caption{\label{xitime}The time dependence of the bond-orientational correlation length $\xi_\Psi$ (solid circles)
and the bond-breaking correlation length $\xi_B$ (open squares)
for $T=1.0, 0.8, 0.7, 0.6,$ and 0.5. Initially, the two correlation lengths are approximately 
equal, but $\xi_\Psi$ stops growing at an earlier time than $\xi_B$. Both length
scales have a period of logarithmic growth (solid line)
for $T \le 0.7$, before a peak and a decrease at later times. The triangles is
the time dependence of the overlap length calculated for $T=0.5$.}
\end{figure}

We begin by examining the time-dependence of the correlation lengths $\xi_B(t)$ and
$\xi_\Psi(t)$. Shown in Fig.~\ref{xitime} are the dynamics correlation lengths $\xi_B(t)$ and $\xi_\Psi(t)$ 
as a function of time for $T=1.0, 0.8, 0.7, 0.6,$ 0.5, and 0.45. For short times, $\xi_B(t)$ (squares) and
$\xi_\Psi(t)$ (circles) are equal and grow approximately as $\ln(t)$ (solid line). 
We note that the $\ln(t)$ growth of the correlation length is similar to what 
was reported for mobility correlations in a three-dimensional hard sphere system \cite{Flenner2011},
and is consistent with the growth of mobility correlations for several 
fragile glass formers \cite{Flenner2014}. 
We found that the growth of $\xi_\Psi(t)$ is arrested before that of $\xi_B(t)$. Both $\xi_\Psi(t)$ 
and $\xi_B(t)$ peak at their specific characteristic times and then decrease for later times.
We recall that in a three-dimensional hard sphere system, there was a plateau in the overlap length and
no peak was observed \cite{Flenner2011}, but a peak in the overlap length 
was reported in the three-dimensional Kob-Andersen system \cite{Shiba2015}. 
While it is difficult to determine the precise location of the peak, $\xi_\Psi(t)$ peaks at a time
approximately five times {\em larger} than $\tau_\Psi$ and $\xi_B(t)$ peaks at a time that is 
about 2.5 times {\em smaller} than $\tau_B$. 

Shown as the triangles in Fig.~\ref{xitime} is the overlap length, \textit{i.e.} 
the length characterizing mobility correlations, $\xi_O(t)$, 
for $T=0.5$. At short times, it grows approximately linearly
with time. This initial linear growth of $\xi_O(t)$ occurs for all temperatures (not shown for clarity). Similar
to the universal growth of $\xi_\Psi(t)$ and $\xi_B(t)$, the time dependence of $\xi_O(t)$ is the same for
each temperature at shorter times and there are deviations from the master curve at later times. 
Importantly, Fig.~\ref{xitime} demonstrates the much stronger growth of the overlap length than the lengths
characterizing spatial correlations of bond-breaking events and bond-orientational relaxation. 

\begin{figure}
\centerline{\includegraphics[width=12cm]{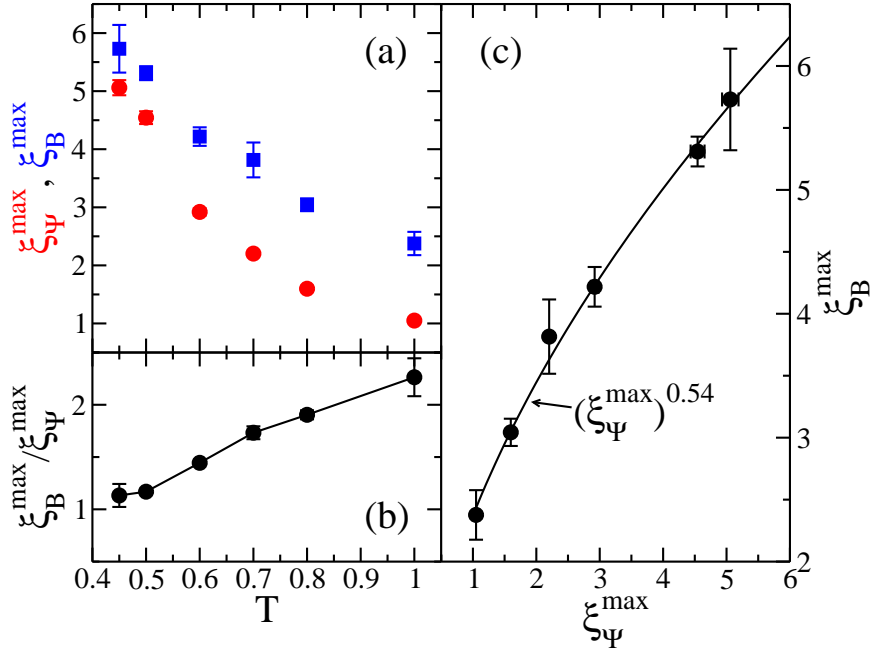}}
\caption{\label{xiratio}(a) The temperature dependence of $\xi_B^{\mathrm{max}}$ and $\xi_\Psi^{\mathrm{max}}$. 
(b) The ratio $\xi_B^{\mathrm{max}}/\xi_\Psi^{\mathrm{max}}$ of
the maximum of the correlation lengths $\xi_B(t)$ and $\xi_\Psi(t)$ as a function
of temperature. 
(c) The correlation length $\xi_B^{\mathrm{max}}$ versus the correlation 
$\xi_\Psi^{\mathrm{max}}$. The line shows a fit to the function 
$\xi_B^{\mathrm{max}} = a (\xi_\Psi^{\mathrm{max}})^\gamma$ with $\gamma = 0.54 \pm 0.05$.}
\end{figure}

To compare the two lengths characterizing spatial correlations of bond-breaking events and bond-orientational relaxation, 
we show the temperature dependence of their maximum values in Fig.~\ref{xiratio}(a). Both lengths 
grow with decreasing temperature, but $\xi_\Psi^{\mathrm{max}}$ (red circles) grows faster than
$\xi_B^{\mathrm{max}}$ (blue squares). In Fig.~\ref{xiratio}(a) we show the ratio of the two lengths, 
$\xi_B^{\mathrm{max}}/\xi_\Psi^{\mathrm{max}}$. We note that the faster 
growth of $\xi_\Psi^{\mathrm{max}}$ is consistent with $\xi_\Psi^{\mathrm{max}}$ occurring at 
a multiple of $\tau_\Psi$, with $\tau_\Psi$ growing faster with decreasing temperature than $\tau_B$.
We note an approximate power-law relationship between the two lengths: a  
fit of $\xi_B^{\mathrm{max}}$ to $a(\xi_\Psi^{\mathrm{max}})^\gamma$ resulting in  
$\gamma = 0.54 \pm 0.05$ represents the data within error, see  Fig.~\ref{xiratio}(c). 
However, a larger range of correlation lengths would be needed to confirm this relationship. In particular,  
it would be interesting to investigate what happens at lower temperatures. 
 
 \section{Dynamic Susceptibilities}
 
 The ensemble independent susceptibility $S_4^w(0;t) = \lim_{q \rightarrow 0} S_4^w(q;t)$ is a
 measure of the strength of the dynamic heterogeneity probed by the
 weight function $w(t)$. However, it can be difficult to 
 calculate $S_4^w(0;t)$ since one has to either simulate large systems \cite{Karmakar2010} or account 
 for suppressed global fluctuations of the simulational ensemble \cite{Berthier2005,Flenner2010,Flenner2011}.  
 Typically, the ensemble dependent susceptibilities are studied and the behavior of the full dynamic
 susceptibility $S_4^w(0;t)$ and the dynamic correlation length $\xi_w(t)$ is inferred. Here
 we will examine the ensemble dependent susceptibilities $\chi_4^w(t)$ and examine how 
 they are related to the ensemble independent susceptibilities and the dynamic correlation lengths.
 
 The ensemble dependent dynamic susceptibilities are given by
 \begin{equation}
 \chi_4^w(t) = \frac{1}{N} \left< \left| \sum_n w_n(t) \right|^2 \right> - \left< \sum_n w_n(t) \right>^2,
 \end{equation}
 where $\left< \cdot \right>$ denotes an average in the $NVT$ ensemble. The ensemble independent susceptibilities 
 $S_4^w(0;t) = \lim_{q \rightarrow 0} S_4^w(q;t)$ are obtained from the fits to $S_4^w(q;t)$. 
 
\begin{figure}
 \centerline{\includegraphics[width=12cm]{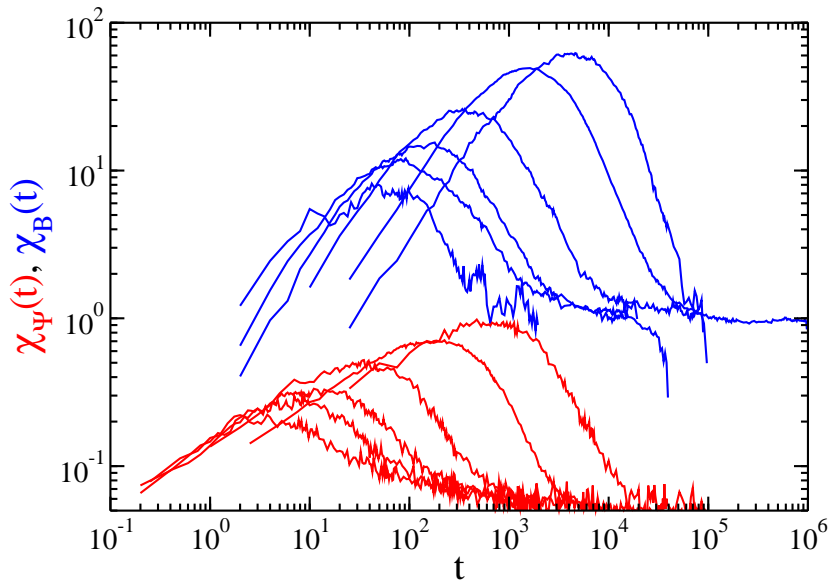}}
 \caption{\label{sus}The ensemble dependent susceptibilities $\chi_4^\Psi(t)$ (lower red lines) and 
 $\chi_4^B(t)$ (upper blue lines) for $T=1.0, 0.8, 0.7, 0.6, 0.5,$ and 0.45 listed from left to right.}
 \end{figure}

\begin{figure}
 \centerline{\includegraphics[width=12cm]{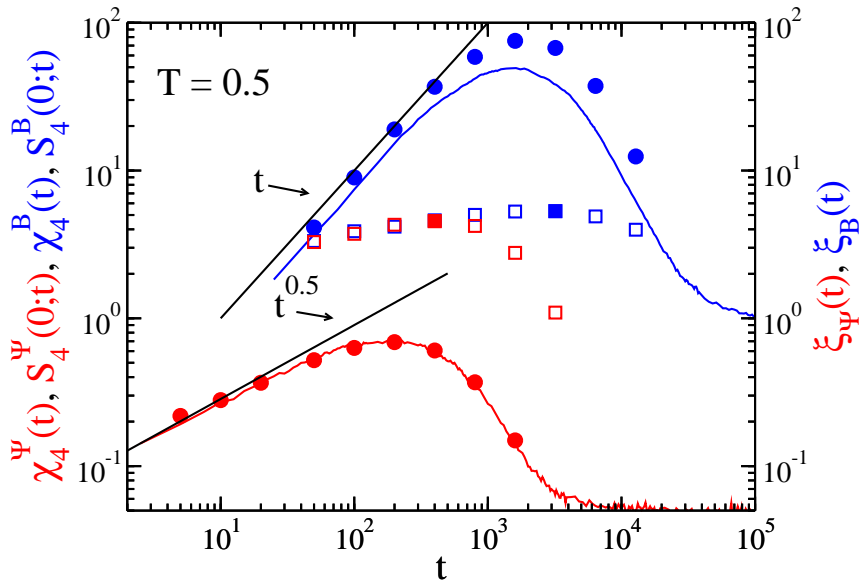}}
 \caption{\label{suscomp}The ensemble dependent susceptibilities $\chi_4^{\Psi}(t)$ (lower red line)
 and $\chi_4^B(t)$ (upper blue line) compared to the ensemble independent susceptibility $S_4^{\Psi}(0;t)$ 
 (lower red circles) and $S_4^B(0;t)$ (upper blue circles) for $T=0.5$. Note that $\chi_4^{\Psi}(t)$ is
 equal to $S_4^{\Psi}(0;t)$ to within error for this temperature. The squares are the 
 correlation lengths $\xi_\Psi(t)$ (red) and $\xi_B(t)$ (blue), where the filled square is the maximum length.
 The maximum length occurs at a time slightly larger than the time of the susceptibility peak.}
 \end{figure}

 Shown in Fig.~\ref{sus} are $\chi_4^\Psi(t)$ (lower red lines) and $\chi_4^B(t)$ (upper blue lines)
 on a log-log scale for $T=1.0, 0.8, 0.7, 0.6, 0.5,$ and 0.45. The bond-orientational susceptibility $\chi_4^\Psi(t)$ grows as 
 $\sqrt{t}$, Fig.~\ref{suscomp}, before it plateaus and then decays to a constant. However
 the bond-breaking susceptibility $\chi_4^B(t)$ grows linearly with time, Fig.~\ref{suscomp}, before the peak. For 
 a given temperature, $\chi_4^B(t)$ peaks at a later time than $\chi_4^\Psi(t)$. 
 
 We compare the ensemble dependent susceptibility $\chi_4^w(t)$ to 
 the ensemble independent susceptibility $S_4^w(0;t)$ in Fig.~\ref{suscomp}. 
 We find that $S_4^B(0;t)$ is larger than $\chi_4^B(t)$, but its peak is at around
 the same time as that of $\chi_4^B(t)$. Note that the ensemble independent susceptibility 
 is expected to be equal to or larger than the ensemble dependent susceptibility \cite{Berthier2005}. 
 Surprisingly, we find that $S_4^\Psi(0;t)$ is equal to $\chi_4^\Psi(t)$ within error
 for $T=0.5$ over the time range we studied. We note that for higher temperatures 
 we found that $S_4^\Psi(0,t)$ is slightly larger than $\chi_4^\Psi(t)$ in general. 
 
 We also show the time dependence of the correlation lengths $\xi_B(t)$ (blue squares)
 and $\xi_\Psi(t)$ (red squares) for $T=0.5$ in Fig.~\ref{suscomp}. As noted in 
 Sec.~\ref{lengths}, $\xi_B(t)$ and $\xi_\Psi(t)$ are initially equal, but $\xi_\Psi(t)$ 
 plateaus and decreases before $\xi_B(t)$. Note that the maximum length, denoted 
 as filled squares, occurs slightly
 after the peak in $\chi_B(t)$ and $\chi_\Psi(t)$ for 
 both correlation lengths. Recall that $\tau_B$ is larger than the time of the maximum of
 $\xi_B(t)$ and $\tau_\Psi$ is smaller than the time of the maximum of $\xi_\Psi(t)$. 
 Therefore, it seems appropriate to compare the correlation lengths around the maximum of the
 susceptibility, which is around the maximum lengths for $\xi_B(t)$ and $\xi_\Psi(t)$. 
 In fact, such a procedure was used in Ref. \cite{Shiba2015}. 

 \section{Conclusions}
 
 We examined bond-orientational relaxation and bond-breaking events in a model 
 two-dimensional glass forming system. We found that the characteristic time for particles
 to lose their neighbors, the bond-breaking relaxation time, is longer than the decay time of 
 bond-orientational correlations and the decay time of the self-intermediate scattering function. 
 However, the bond-orientational correlation time increases faster with decreasing 
 temperature than the bond-breaking relaxation time over the temperature range studied. 
 
 We also examined the spatial correlations of bond-orientational relaxation and bond-breaking events.
 We found that at each temperature the characteristic size of correlated domains of bond-orientational 
 relaxation and bond-breaking events were equal at short times, but the characteristic size
 of domains of correlated bond-orientational relaxation saturates at an earlier time than for 
 bond-breaking events and the associated maximum correlation length is smaller.  However,
 the maximum correlation length for bond-orientational relaxation grows faster with 
 decreasing temperature than the maximum correlation length for bond-breaking events.
 The correlation lengths appear to be related by a power law where 
 $\xi_B^{\mathrm{max}} \propto (\xi_\Psi^{\mathrm{max}})^\gamma$
 with $\gamma \approx 0.54$, but a larger range of length scales would needed to verify this
 relationship. 
 
 While we did not verify that the correlations of bond-orientational relaxation grow as
 mobility correlations in three dimensions, this result would be expected since bond-orientational
 relaxation time grows as structural relaxation time in three dimensions. It has already 
 been demonstrated that bond-breaking correlation length grows as mobility correlation 
 length in three-dimensions, 
 but bond-breaking correlation length grows slower than
 mobility correlation length in two dimensions \cite{Shiba2015}. However, in two dimensions there are now
 three time dependent correlation lengths with different temperature dependence. We note
 that the bond-orientational and the bond-breaking correlation lengths appear to be related, 
 and future studies of lower temperature two-dimensional systems needs to be performed to
 understand this relationship fuller. 
 
 We found that the ensemble dependent susceptibility calculated in an NVT ensemble 
 and the ensemble independent susceptibility each 
 has a peak at a time slightly less than the time of the maximum of the 
 bond-orientational and the bond-breaking correlation lengths. Importantly,
 the ensemble independent and the ensemble dependent susceptibilities peak at the same time. Note that the
 peak time of the bond-orientational dynamic susceptibility is larger than the bond-orientational relaxation 
 time, but the peak time of the bond-breaking dynamic susceptibility is at a time smaller 
 than the bond-breaking relaxation time. Therefore, the dynamic correlation length calculated at
 the bond-orientational relaxation time is still growing, but the dynamic correlation length calculated
 at the bond-breaking relaxation time has already reached its peak and is decreasing.
 
 \ack
 We gratefully acknowledge funding from NSF Grant No.~CHE 1213401. 
 
 \appendix
 \section{Calculating Dynamic Correlation Lengths}
 \begin{figure}
 \centerline{\includegraphics[width=12cm]{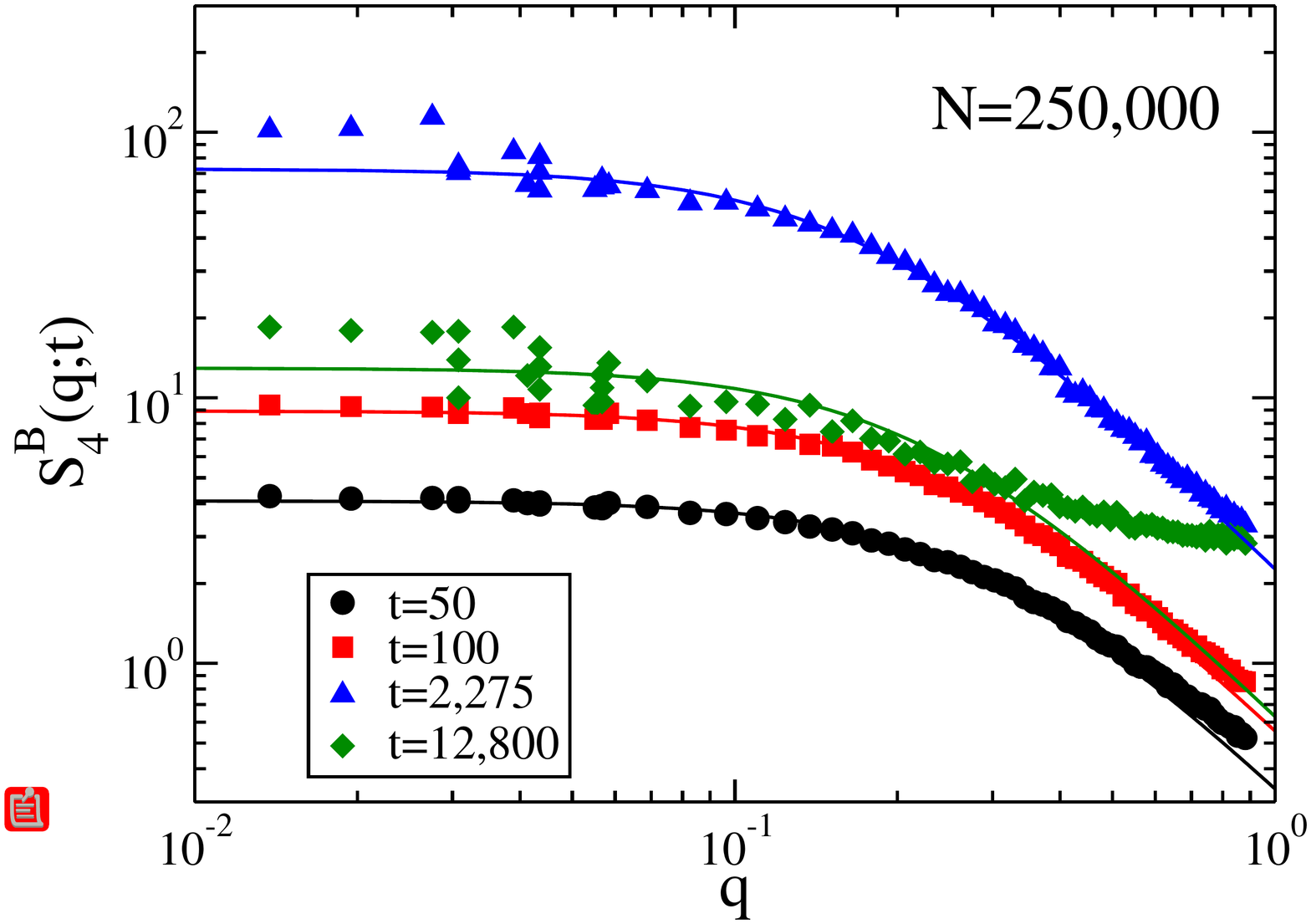}}
 \caption{\label{S4bb}Some example four-point structure factors $S_4^B(q;t)$ 
 for $T=0.5$. For $q \approx 0.3$ and $q \approx 0.04$ several wave-vectors with the same magnitude but
 are calculated with $q$ in different directions are shown. The lines are Ornstein-Zernicke fits. The range of 
 $q$ values are chosen such that the Ornstein-Zernicke fits give a correlation length that is 
 consistent with analyzing $G_4^B(r;t)$.}
 \end{figure}
 \begin{figure}
 \centerline{\includegraphics[width=12cm]{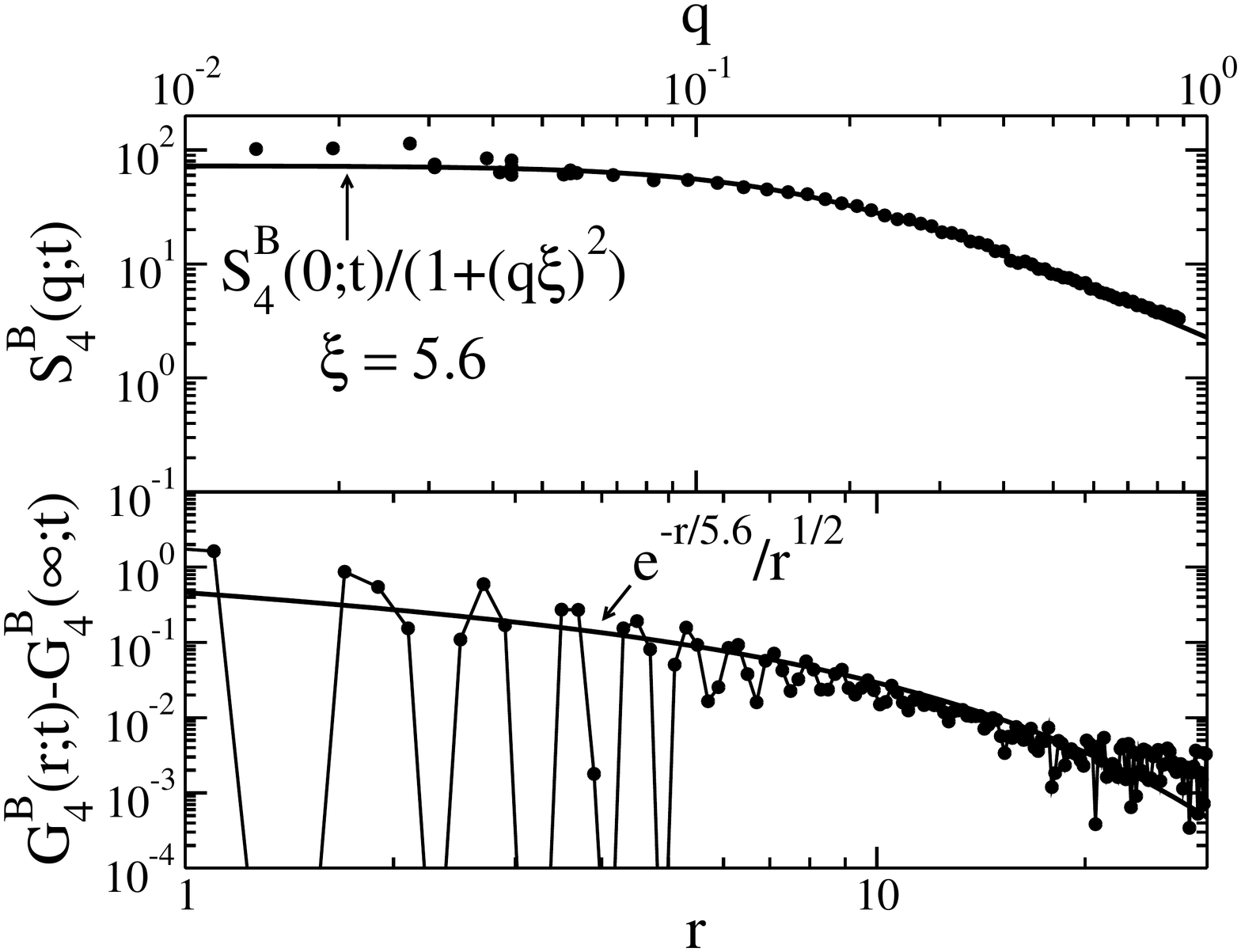}}
 \caption{\label{example}The upper figure shows the four-point structure factor $S_4^B(q;t)$ and the 
 Ornstein-Zernicke fit to obtain the correlation length at $t=2,275$,
 which is close to time of the peak of the dynamic susceptibility $\chi_4^B(t)$. The lower 
 figure is the associated four-point correlation function $G_4^B(r;t) - G_4^B(\infty;t)$
 and the verification that the length from the Ornstein-Zernicke fit is compatible with 
 the decay of $G_4^B(r;t)$.}
 \end{figure}

 In this appendix we outline the procedure we use to obtain dynamic correlation lengths in our study. 
 We found that it was difficult to determine how to 
 fit the small wave-vector behavior of $S_4^w(q;t)$ from the four-point structure factor alone at
 times later than the peak of the dynamic susceptibility due to noise for small wave-vectors. 
 
 Shown in Fig.~\ref{S4bb} is $S_4^B(q;t)$ at T=0.5 for several times. For short times, the
 small $q$ values converge and the Ornstein-Zernicke function provided good fits 
 to $S_4^B(q;t)$. Several fits are shown in the figure. At later times, there is 
 significant noise in the small $q$ values of $S_4^B(q;t)$ and fits over different
 ranges of $q$ resulted in a large range of correlation lengths. To determine the correct
 correlation length, we also calculated the four-point pair correlation function
 \begin{equation}
 G_4^w(r;t) = \frac{V}{N^2} 
 \left< \sum_n \sum_{m \ne n} w_n(t) w_m^*(t) \delta(\mathbf{r} - \mathbf{r}_{nm}(0)) \right>, 
 \end{equation}
 where $\mathbf{r}_{nm}(0) = \mathbf{r}_n(0) - \mathbf{r}_m(0) $. We checked
 that the large $r$ decay of $G_4^w(r;t)$ was consistent with the Ornstein-Zernicke fit. 
 To this end we found $G_4^w(r;t)$ for large $r$ by fitting $G_4^w(r;t)$ for 
 $r > 100$ to a constant $G_4^w(\infty,t)$. We then examined $G_4^w(r;t) - G_4^w(\infty,t)$
 and verified that it was consistent with the Ornstein-Zernicke fit by verifying that the 
 decay followed $e^{-r/\xi}/\sqrt{r}$. This functional form is obtained from noting
 that the inverse Fourier Transform in two dimensions of $1/[1+(q\xi)^2]$ is proportional
 to $K_0(r/\xi)$ where $K_0$ is the modified Bessel function of the second kind, 
 and that $K_0(r/\xi)$ 
 is well approximated by $e^{-r/\xi}/\sqrt{r}$ for large $r$. We show an example
 of this verification for $T=0.5$ and $t=2\, 275$ in Fig.~\ref{example}. To 
 calculate $G_4^B(r;t)$ for large systems we used the technique introduced 
 by Bernard and Krauth \cite{Bernard2011}.

\end{document}